\documentclass[a4paper,12pt]{article}
\begin{document}

\title{Impossibility of Exact Flipping of Three Arbitrary Quantum States Via Incomparability}

\author{Indrani Chattopadhyay\thanks{ichattopadhyay@yahoo.co.in}\ \ and Debasis Sarkar
\thanks{dsappmath@caluniv.ac.in, debasis1x@yahoo.co.in}}

\maketitle

\begin{center}
Department of Applied Mathematics, University of Calcutta,\\ 92,
A.P.C. Road, Kolkata- 700009, India.
\end{center}

\begin{abstract}
We present here a scheme that relates seemingly two different kinds
of physical impossibilities of quantum information processing. We
derive, exact flipping of three arbitrary states not lying in one
great circle is not possible with certainty, by using the existence
of incomparable states which are not interconvertible
deterministically by local operations and classical communications.
In contrast, considering the non-existence of exact universal
flipper, the incomparability of a pair of bipartite pure entangled
states can be established.

PACS number(s): 03.67.Hk, 03.65.Ud.

\end{abstract}
\maketitle \vspace{.5cm}

The linear superposition principle of quantum mechanical amplitudes
give it such a strange structure that some operations which are
obvious in classical domain are found to be impossible in the
quantum arena. Some well known impossibilities are, no-cloning
\cite{wootters}, no-deleting \cite{pati11}, no-flipping
\cite{gisin}, etc., \cite{patili}. Efforts are made to find
inter-relations between those impossibilities and the basic dynamics
of quantum mechanics. For example, the no-cloning principle directly
follows from the linearity and unitary dynamics of the quantum
mechanical systems {\cite {linear}}. In spite of the non-local
character of quantum mechanics \cite{mes}, it cannot be used to send
signals faster than light. We find a peaceful coexistence between
no-signalling condition and the no-go principles, like, no-cloning,
no-deleting, probabilistic cloning, no-flipping, etc., \cite{signal,
flip}.

The structure of the allowed operations performed on the quantum
systems shared between some spatially separated parties imposes
some restrictions on the systems. Allowing local operations along
with classical communications (in short, LOCC) between separated
parties, one could find some impossibilities, like the principle
of `Non-increase of entanglement by LOCC'. This has been connected
with no-cloning, no-deleting \cite{horo} and further with the
no-flipping principle \cite{flip}. In a quite similar manner we
could find another impossibility from the transformation of pure
bipartite entangled states which we may frame as `Incomparable
states can not be deterministically transformed into one another
by LOCC performed on the subsystems' \cite{nielsen,
incomparability}. The peculiarity of the existence of incomparable
states is that it doesn't care about the amount of entanglement
contained in the states. Though one of the two states has a
greater entanglement content than the other, but the
transformation can not be happen by deterministic LOCC for a pair
of incomparable states. The existence of such class of states
itself shows a peculiar feature of LOCC which can be studied to
explore other areas of restrictions on the physical systems. Here
we explore its connection with an impossible operation of quantum
system.

Non-existence of universal exact flipping machine is a kind of
constraint on the quantum systems that has been directly observed
from unitary dynamics of the quantum evolution \cite{gisin}. One
could verify that the universal exact flipper, which if operated
on an arbitrary state, will reverse the spin polarization
direction, is not unitary but an anti-unitary operation and thus
it is not a physical operation. No-flipping principle says that
exact flipping of an unknown qubit state, is not possible. Like
no-cloning, no-deleting principles, it is also a fundamental
restriction to the allowable operations on a quantum system though
this is an impossibility acting only on a single quantum system
whereas we require at least two subsystems to describe no-cloning,
no-deleting principles. One interesting observation is found by
Gisin and Popescu \cite{gisin} and also by Massar \cite{massar}
that the two-anti parallel-spin state
$|\overrightarrow{n},\overrightarrow{-n}\rangle$ contains more
information than two-parallel-spin state
$|\overrightarrow{n},\overrightarrow{n}\rangle.$ It is conjectured
that the origin of this feature is the property that the anti
parallel spin states span the whole four dimensional Hilbert space
of two spin $\frac{1}{2}$ system, while the parallel spin
$\frac{1}{2}$ system spans only a three dimensional subspace of
the symmetric states. Then the optimal flipping machine is
proposed by Buzek et.al. \cite{unot}. The no-flipping principle
has a completely different status from that of no-cloning,
no-deleting principles as unlike those cases, there always exist a
quantum machine which can act as an exact flipper for any two
non-orthogonal qubit states. Not only that, the largest set of
states that can be flipped by a single flipper machine is any
great circle of the Bloch sphere \cite{ghosh}. It is also found
that any three states of the Bloch sphere, not lying in one great
circle can not be flipped by a single quantum machine. This
no-flipping principle has already shown to be a consequence of
no-signalling and non increase of entanglement by local operations
\cite{flip}.

Here we present a scheme that relates the impossibility of exact
flipping of arbitrary qubit states with a seemingly different kind
of impossibility of inter conversion of incomparable states by
deterministic LOCC assuming the usual quantum mechanical
formalism. We first show the impossibility of exact flipping for
the three Bloch vectors along $x, y, z,$ through incomparability.
Then we show that by assuming the inter conversion of two
incomparable states by deterministic LOCC one can generate a
flipping machine for the whole Bloch sphere that reverses the
directions of every Bloch vectors. Lastly we consider the first
problem, the impossibility of universal exact flipping machine for
any three qubits not lying on a great circle via incomparability.
Now before going to describe our results, we first briefly mention
incomparability criterion of a pair of pure bipartite entangled
states.

The notion of incomparability of a pair of bipartite pure states
is prescribed by M. A. Nielsen \cite{nielsen}. Suppose we want to
convert the pure bipartite state $|\Psi\rangle$ to $|\Phi\rangle$
shared between two parties, say, Alice and Bob by deterministic
LOCC. Consider $|\Psi\rangle$, $|\Phi\rangle$ in the Schmidt bases
$\{|i_A\rangle ,|i_B\rangle \}$ with decreasing order of Schmidt
coefficients: $|\Psi\rangle= \sum_{i=1}^{d} \sqrt{\alpha_{i}} |i_A
i_B\rangle$, $|\Phi\rangle= \sum_{i=1}^{d} \sqrt{\beta_{i}} |i_A
i_B\rangle,$ where $\alpha_{i}\geq \alpha_{i+1}\geq 0$ and
$\beta_{i}\geq \beta_{i+1}\geq0,$ for $i=1,2,\cdots,d-1,$ and
$\sum_{i=1}^{d} \alpha_{i} = 1 = \sum_{i=1}^{d} \beta_{i}$. And
denote the Schmidt vectors as
$\lambda_\Psi\equiv(\alpha_1,\alpha_2,\cdots,\alpha_d),$
$\lambda_\Phi\equiv(\beta_1,\beta_2,\cdots,\beta_d)$. Then
Nielsen's criterion says $|\Psi\rangle\rightarrow| \Phi\rangle$ is
possible with certainty under LOCC if and only if $\lambda_\Psi$
is majorized by $\lambda_\Phi,$ denoted by
$\lambda_\Psi\prec\lambda_\Phi$ and described as,
\begin{equation}
\begin{array}{lcl}\sum_{i=1}^{k}\alpha_{i}\leq
\sum_{i=1}^{k}\beta_{i}~ ~\forall~ ~k=1,2,\cdots,d
\end{array}
\end{equation}
As a consequence of non-increase of entanglement by LOCC, if
$|\Psi\rangle\rightarrow |\Phi\rangle$ is possible under LOCC with
certainty, then $E(|\Psi\rangle)\geq E(|\Phi\rangle)$ [where
$E(\cdot)$ denote the von-Neumann entropy of the reduced density
operator of any subsystem and known as the entropy of
entanglement]. If the above criterion (1) does not hold, then it
is usually denoted by $|\Psi\rangle\not\rightarrow |\Phi\rangle$.
Though it may happen that $|\Phi\rangle\rightarrow |\Psi\rangle$
under LOCC. Now if it happens that $|\Psi\rangle\not\rightarrow
|\Phi\rangle$ and $|\Phi\rangle\not\rightarrow |\Psi\rangle$ then
we denote it as $|\Psi\rangle\not\leftrightarrow |\Phi\rangle$ and
describe $(|\Psi\rangle, |\Phi\rangle)$ as a pair of incomparable
states. One of the interesting feature of such incomparable pairs
is that we are unable to say that which state has a greater amount
of entanglement content than the other. Also for $2\times 2$
systems there are no pair of pure entangled states which are
incomparable as described above. Now we explicitly mention the
criterion of incomparability for a pair of pure entangled states
$|\Psi\rangle, |\Phi\rangle$ of $m\times n$ system where $\min \{
m,n \}=3$. Suppose the Schmidt vectors corresponding to the two
states are $(a_1, a_2, a_3)$ and $(b_1, b_2, b_3)$ respectively,
where $a_1> a_2> a_3~,~b_1> b_2> b_3~,~a_1+ a_2+ a_3=1=b_1+ b_2+
b_3$. Then it follows from Nielsen's criterion that $|\Psi\rangle,
|\Phi\rangle$ are incomparable if and only if either of the two
relations,
\begin{equation}
\begin{array}{lcl}
a_1 > b_1 ~ ~\verb"and" ~ ~b_1 + b_2 > a_1 + a_2\\ b_1 > a_1 ~
~\verb"and" ~ ~
a_1 + a_2 > b_1 + b_2 ~ ~ \verb"hold,"\\
\verb"or alternately,"\\

a_1 > b_1 >  b_2 > a_2 > a_3 > b_3\\ b_1 > a_1 >  a_2 > b_2
> b_3 > a_3
\end{array}
\end{equation} hold.

At first to sense the idea of our work, consider three states
representing the three axis of the Bloch sphere in the usual
basis, as $|\psi_x
\rangle=\frac{|0\rangle+|1\rangle}{\sqrt{2}}~,~|\psi_y
\rangle=\frac{|0\rangle+i|1\rangle}{\sqrt{2}}~,
|\psi_z\rangle=|0\rangle$. We choose the particular setting of a
pure bipartite state in the form
\begin{equation}
\begin{array}{lcl}
|\Psi^i\rangle_{AB}=
\frac{1}{\sqrt{3}}~\{~|0\rangle_A|\psi_z\psi_z\rangle_B +
|1\rangle_A|\psi_x\psi_y\rangle_B +
|2\rangle_A|\psi_y\psi_x\rangle_B~\}
\end{array}
\end{equation}
This is a three particle state, shared between two space-like
separated parties Alice and Bob, so that Alice has a qutrit and
Bob has two qubits, i.e., as a bipartite state it belongs to a
$3\times 4$ system. The reduced density matrix of Bob's subsystem
admits a representation in terms of the three states $|\psi_x
\rangle, |\psi_y \rangle, |\psi_z \rangle$. The Schmidt vector
corresponding to the reduced system on Alice's side is$\lambda^i
=(\frac{2}{3},\frac{1}{6},\frac{1}{6})$ . If the existence of
exact flipping machine for the three states $~|\psi_x
\rangle,~|\psi_y \rangle,~|\psi_z \rangle$ is possible, then by
applying this machine to one of the two particles on Bob's side,
the joint state between them can be exactly transformed to the
state
\begin{equation}
\begin{array}{lcl}
|\Psi^f
\rangle_{AB}=\frac{1}{\sqrt{3}}~\{~|0\rangle_A|\psi_z\overline{\psi_z}~\rangle_B
+ e^{i\chi} |1\rangle_A|\psi_x\overline{\psi_y}~\rangle_B +
e^{i\eta}|2\rangle_A|\psi_y\overline{\psi_x}~\rangle_B~\}
\end{array}
\end{equation}
where $\chi,\eta$ are some arbitrary phases. The final reduced
density matrix on Alice's side has the Schmidt vector
$\lambda^f=(\frac{1}{3}+\frac{1}{2\sqrt{3}}~, \frac{1}{3}~,
\frac{1}{3}-\frac{1}{2\sqrt{3}})$. From equation (2) it is easy to
check that $|\Psi^i \rangle_{AB},~|\Psi^f\rangle_{AB}$ is a pair
of incomparable states. Hence by Nielsen's criterion it is
impossible to locally transform $|\Psi^i \rangle_{AB}$ to $|\Psi^f
\rangle_{AB}$ with certainty. In this example we see that the
impossibility of an operation in quantum mechanics can be
established from the contradiction that it forces two incomparable
states to become comparable by LOCC with certainty.

The result shows, how a local anti-unitary operation evolves the
system in an unphysical way. To explore this unphysical nature of
anti-linear operators, we search for a joint system between some
separated parties, as on a single system it is really difficult to
distinguish unitary and anti-unitary operators. Here we observe
that applying an anti-unitary operator (say, $L$) locally on a
joint system of $3 \times (2 \times 2)$ dimension, i.e., applying
the operator $I_A \otimes (I \otimes L)_B$ on the joint system,
there arises a case of incomparability. The bi-partite
entanglement of the joint system changes and it changes in such a
manner that there is no way to compare the initial and final joint
states locally. Now one can ask whether the converse is also to be
happen or not, i.e., is it possible to generate an exact flipping
machine by assuming inter conversion of at least a pair of
incomparable states by deterministic LOCC? Our next example is
going to explore it where by exact flipping we mean a machine that
can reverse the direction of any Bloch vector. This machine
usually generalizes the notion of flipping from any pure qubit to
any state on the whole Bloch sphere.

Consider a pair of pure entangled states, shared between Alice and
Bob situated in two distant places, in the form
\begin{equation}
\begin{array}{lcl}
|\Psi\rangle_{AB}= \sqrt{.51}~|0\rangle_A|0\rangle_B +
\sqrt{.30}~|1\rangle_A|1\rangle_B +
\sqrt{.19}~|2\rangle_A|2\rangle_B,\\
|\Phi\rangle_{AB}= \sqrt{.49}~|0\rangle_A|0\rangle_B +
\sqrt{.36}~|1\rangle_A|1\rangle_B +
\sqrt{.15}~|2\rangle_A|2\rangle_B.
\end{array}
\end{equation}
The Schmidt vectors corresponding to one of the subsystems of $|\Psi
\rangle_{AB}, |\Phi \rangle_{AB}$ are $\lambda^{\Psi}=
(.51,.30,.19)$ and $\lambda^{\Phi}=(.49,.36,.15)$. Using equation
(2) we find, $(|\Psi \rangle_{AB}, |\Phi \rangle_{AB})$ are a pair
of incomparable states. Suppose Bob has a two qubit system on his
side and the orthogonal states $\{|0\rangle_B, |1\rangle_B,
|2\rangle_B \}$ have the form,

$|0\rangle_B=|\psi\rangle_{B_1}|\psi\rangle_{B_2}~,~|1\rangle_B=
|\overline{\psi}\rangle_{B_1}|\psi\rangle_{B_2}~,
~|2\rangle_B=|\overline{\psi}\rangle_{B_1}|\overline{\psi}\rangle_{B_2}~,$\\
where $|\psi \rangle$ is an arbitrary qubit state with Bloch
vector $\overrightarrow{n_{\psi}}$, i.e., $|\psi \rangle \langle
\psi | = \frac{1}{2}[I + \overrightarrow{n_{\psi}}\cdot
\overrightarrow{\sigma}]$ and $|\overline{\psi}\rangle$ is
orthogonal to $|\psi \rangle$.

Now tracing out Alice's system and the second qubit of Bob's side
(i.e., system $B_2$), we get one qubit reduced subsystem of Bob
corresponding to the two joint systems in (5) as,
\begin{equation}
\begin{array}{lcl}
\rho^{\Psi}= Tr_{A B_2 }\{| \Psi \rangle \langle \Psi |\}=~
\frac{1}{2}~[I + .02~\overrightarrow{n_{\psi}}\cdot
\overrightarrow{\sigma}]\\
\rho^{\Phi}= Tr_{A B_2 }\{| \Phi \rangle \langle \Phi |\}=~
\frac{1}{2}~[I - .02~\overrightarrow{n_{\psi}}\cdot
\overrightarrow{\sigma}].
\end{array}
\end{equation}
If by any local operation it is possible to transform the joint
state between Alice and Bob from $|\Psi\rangle$ to $|\Phi\rangle$,
then the reduced state of one qubit subsystem on Bob's side will
be changed accordingly from $\rho^{\Psi}$ to $\rho^{\Phi}$
exactly, by LOCC. It is clear that the spin direction
$\overrightarrow{n_{\psi}}$ of the arbitrary qubit state
$|\psi\rangle$ is reversed after the operation, i.e., transformed
to $-\overrightarrow{n_{\psi}}$. So, if we extend the LOCC
transformation criterion so that the states $|\Psi\rangle,
|\Phi\rangle$ are interconvertible by some operation then
consequently on a subsystem, the spin-polarization of an arbitrary
qubit state is being reversed. This is quite similar of preparing
an arbitrary spin flipper machine and is an alternative way of
establishing the incomparable nature of the pair of states in
equation (5).

This two examples in contrast of each other show, how an
impossible local operation is connected with the restrictions
imposed on state transformation by LOCC. The root of this
connection is the anti-unitary nature of the exact universal
spin-flipping operation. In the case of state transformation
criterion, the allowed local operations on each parties are such
that as a whole it can be implemented by an unitary evolution and
if we restrict further individual local operations as unitary then
bipartite entanglement cannot be changed under such local unitary
operations. We, however consider a local operation which is
anti-unitary and it is observed that anti-unitary operator acts in
a nonphysical way. The above two examples show that it is not
possible to extend the class of physical operations so as to
incorporate anti-unitary operations \cite{gisin1}.

To generalize the main result corresponding to the first example,
we consider three arbitrary states not lying in one great circle
in their simplest form, $|0\rangle,~ |\psi\rangle= a|0\rangle +
b|1\rangle, ~|\phi\rangle= c|0\rangle + d~e^{i\theta}|1\rangle$
where a, b, c, d are real numbers satisfying the relation $ a^2 +
b^2 =~ 1=~ c^2 + d^2~$ and $0 < \theta < \pi $. Suppose two
spatially separated parties Alice and Bob are sharing the
entangled state
\begin{equation}
\begin{array}{lcl}
|\Omega\rangle_{AB}= \frac{1}{\sqrt{3}}~\{~|0\rangle_A|00\rangle_B
+|1\rangle_A|\psi\phi\rangle_B +|2\rangle_A|\phi\psi\rangle_B~\}
\end{array}
\end{equation}
where Alice has a 3-dimensional orthogonal local system, having
the basis, $\{|0\rangle,~|1\rangle,~|2\rangle \}$ and Bob has a
two qubit system. The reduced density matrix of Alice's side is
\begin{equation}
\begin{array}{lcl}
\rho_A^i &=&\frac{1}{3}~ \{P[|0\rangle]+P[|1\rangle]+P[|2\rangle]+
ac(|0\rangle\langle1|+|1\rangle\langle0|+|0\rangle\langle2|
\\ & &~~+|2\rangle\langle0|)+|\langle \psi |
\phi\rangle|^2(|1\rangle\langle2| +|2\rangle\langle1|) \}
\end{array}
\end{equation}
Consider that Bob has another system which acts as a exact
flipping machine defined on these three states
$|0\rangle,~|\psi\rangle,~|\phi\rangle$. The flipping operation
can be described as
\begin{equation}
\begin{array}{lcl}
& & |0\rangle \longrightarrow |1\rangle  \\
& & |\psi\rangle \longrightarrow
e^{i\mu}|\overline{\psi}\rangle  \\
& &  |\phi\rangle \longrightarrow e^{i\nu}|\overline{\phi}\rangle
\end{array}
\end{equation}
where  $|\overline{\psi} \rangle,| \overline{\phi} \rangle$ are
the states orthogonal to $|\psi\rangle,~|\phi\rangle$ respectively
and $\mu$ and $\nu$ are some arbitrary phases. Now assume that Bob
applies the flipping machine on any one of his two particles (say,
on the second particle). After this local operation on Bob's
subsystem, the shared state between Alice and Bob takes the form,
\begin{equation}
\begin{array}{lcl}|\Omega\rangle^{f}_{AB}~=~\frac{1}{\sqrt{3}}~ \{|0\rangle_A |01\rangle_B +
e^{i \nu}~|1\rangle_A|\psi\overline{\phi}\rangle_B + e^{i
\mu}~|2\rangle_A|\phi\overline{\psi}\rangle_B \}
\end{array}
\end{equation}
The final density matrix of Alice's side is
\begin{equation}
\begin{array}{lcl}
\rho^{f}_A &=&\frac{1}{3}~ \{~P[|0\rangle] +
P[|1\rangle]+P[|2\rangle]\\
& &- ac(~e^{-i\nu} |0 \rangle\langle 1| + e^{i\nu}
|1\rangle\langle 0|+e^{-i\mu} |0\rangle\langle2|  + e^{i\mu}
|2\rangle\langle0|)\\
& & +(\langle \phi | \psi\rangle )^2 e^{i(\nu-\mu)} |1\rangle
\langle 2| +(\langle \psi | \phi\rangle )^2 e^{i(\mu-\nu)}
|2\rangle \langle 1|  ~\}
\end{array}
\end{equation}
The eigenvalue equation on the initial local density matrix
$\rho^i_A$ is,
\begin{equation}
\begin{array}{lcl}
(1-3\lambda)^3 - 3(1-3\lambda)A+ B=0
\end{array}
\end{equation}
and that of the final local density matrix $\rho^{f}_A$ is,
\begin{equation}
\begin{array}{lcl}
(1-3\lambda)^3 - 3(1-3\lambda)A + B'=0
\end{array}
\end{equation}
where $A=\frac{1}{3}[2a^2c^2+{|\langle \psi | \phi\rangle
|}^4],~B=2a^2c^2{|\langle \psi | \phi\rangle |}^2$ and $
B'=2a^2c^2~ Re\{{\langle \phi | \psi\rangle }^2\}.$ It is to be
noticed that the phase factors $e^{i\mu}, e^{i\nu}$ vanishes from
this stage. So the result obtained doesn't care the phase factor
of the operation. The roots of the equation (12) are
$\alpha_1=\frac{1}{3}\{1 - 2\sqrt{A}\cos
(\frac{2\pi}{3}+\theta^i)\}, ~\alpha_2= \frac{1}{3}\{1 - 2\sqrt{A}
\cos {\theta^i}\}$ and $\alpha_3= \frac{1}{3}\{1 - 2\sqrt{A}
\cos(\frac{2\pi}{3}-\theta^i)\}$ and the roots of the equation
(13) are $\beta_1= \frac{1}{3}\{1 -2\sqrt{A}\cos
(\frac{2\pi}{3}+\theta^f)\}, ~\beta_2= \frac{1}{3}\{1 - 2\sqrt{A}
\cos {\theta^f}\}$ and $\beta_3= \frac{1}{3}\{1 - 2\sqrt{A}
\cos(\frac{2\pi}{3}-\theta^f)\}$ where,
$\cos{3\theta^i}~=~\frac{-B}{2\sqrt{A^3}}$, and
$\cos{3\theta^f}~=~\frac{-B'}{2\sqrt{A^3}}$. Now to compare the
states $|\Omega\rangle_{AB}$ to $|\Omega\rangle^f_{AB}$ we do not
need the explicit values of the Schmidt coefficients of the
corresponding reduced density matrices but only we have to find
the relation between them.

Here $B=B'+4a^2 b^2 c^2 d^2 \sin^2 \theta$, so $B\geq 0, B\geq
B'$. If, $0<B'<B,$ then we have
$0>\cos(3\theta^f)>\cos(3\theta^i)$, this will imply that
$3\theta^i, 3\theta^f \in (\frac{\pi}{2},\frac{3\pi}{2})$. We find
four subcases corresponding to the different regions of $\theta^i,
\theta^f$. Firstly we consider the case when $3\theta^i, 3\theta^f
\in (\frac{\pi}{2},\pi).$ In this region, $\cos(3\theta^f)>
\cos(3\theta^i) \Rightarrow 3\theta^i > 3\theta^f$. So, we have
$\frac{\pi}{6}< \theta^f< \theta^i< \frac{\pi}{3},$ which imply
$\frac{\sqrt{3}}{2}>\cos(\theta^f)> \cos(\theta^i)>\frac{1}{2}$.
Then $\alpha_2,\beta_2$ will lie on the region,
$\frac{1}{3}(1-\sqrt{A})> \alpha_2> \beta_2
> \frac{1}{3}(1-\sqrt{3A})$. Proceeding in this way, we find
$~\frac{1}{3}(1+2\sqrt{A})>\alpha_1>\beta_1>\frac{1}{3}(1+\sqrt{3A})$
and $\frac{1}{3}>\beta_3>\alpha_3>\frac{1}{3}(1-\sqrt{A})$. Thus
the eigenvalues of $\rho^i_A,~\rho^{f}_A$ are related as $~
~\alpha_1>\beta_1>\beta_3>\alpha_3>\alpha_2>\beta_2.$ So by
equation (2) the states $|\Omega\rangle_{AB},$
$|\Omega\rangle^f_{AB}$ are incomparable in this region.

In a similar manner we investigated the other regions (for cases
$B'>0$ and $B'<0$) which also show incomparability between the two
bipartite states (See Appendix).

Equations (12) and (13) will be identical (and hence the Schmidt
vectors of $\rho^i_A$ and $\rho^{f}_A$) when $B=B'$, which imply
$abcd \sin\theta=0$, i.e., the three states $|0\rangle,
|\psi\rangle, |\phi\rangle$, on which the flipping machine is
defined, will lie on one great circle of the Bloch sphere. This is
clear from the fact that there exists exact flipping machine for
the set of states taken from one great circle \cite{flip, ghosh}.

Thus, if the exact flipping machine does exist, and is applied
locally on one subsystem of the initial pure bipartite state, then
an impossible transformation is shown to occur. Obviously this
impossibility comes through our assumption on the existence of
universal exact flipping machine. It is interesting to observe
that the arbitrary phase factor of the flipping operation we have
considered, does not make a difference in the result obtained.

This work shows an interplay between the notion of incomparability
and no-flipping principle. It indicates no-flipping can be used to
determine the interrelations between LOCC and entanglement
behavior of the quantum system. We observe, the incomparability
criterion of local state transformations is also capable of
revealing some more fundamental properties of the quantum systems.
It can detect operations which are nonphysical in nature such as,
here it is anti-unitary. Naturally one could conjecture that the
two impossibilities are equivalent, as they both require
anti-unitary operators. Our results support this conjecture. It
also exhibit the impossibility of extending LOCC operations to
incorporate anti-unitary operators which can create an increase of
information content of the system, as anti-parallel spin states
contain more information than that of the parallel ones.

{\bf Acknowledgement.} The authors thank the referees for their
valuable comments and suggestions. The authors also thank G. Kar
for useful discussions in preparation of this paper. I.C.
acknowledges CSIR, India for providing fellowship during this
work.

\begin{center}
\emph{\textbf{ Appendix}}
\end{center}
For $0<B'<B,$ we have the other three possible regions as
follows.\\
When both of $3\theta^i, 3\theta^f~\in (\pi,\frac{3\pi}{2})$ then,
$\alpha_1>\beta_1>\beta_2>\alpha_2>\alpha_3>\beta_3.$\\
When, $3\theta^i\in(\frac{\pi}{2},\pi)$ and $3\theta^f\in
(\pi,\frac{3\pi}{2})$ then,
$\alpha_1>\beta_1>\beta_2>\alpha_3>\alpha_2>\beta_3$.\\
When, $3\theta^i\in(\pi,\frac{3\pi}{2})$ and $3\theta^f\in
(\frac{\pi}{2},\pi)$ then,
$\alpha_1>\beta_1>\beta_3>\alpha_2>\alpha_3>\beta_2.$

Otherwise, $B'<0<B,$ then we have
$\cos(3\theta^f)>0>\cos(3\theta^i)$, which implies $3\theta^i~\in
(\frac{\pi}{2},\frac{3\pi}{2})$ and $3\theta^f~\in
\{(0,\frac{\pi}{2})\bigcup(\frac{3\pi}{2},2\pi)\}$. The following
subcases for different regions of $\theta^i, \theta^f$
are considered separately.\\
When, $3\theta^i\in (\frac{\pi}{2},\pi)$ and $3\theta^f\in
(0,\frac{\pi}{2})$ then,
$\alpha_1>\beta_1>\beta_3>\alpha_3>\alpha_2>\beta_2.$\\
When, $3\theta^i\in (\frac{\pi}{2},\pi)$ and $3\theta^f\in
(\frac{3\pi}{2},2\pi)$ then,
$\alpha_1>\beta_1>\beta_2>\alpha_3>\alpha_2>\beta_3.$\\
When, $3\theta^i\in(\pi,\frac{3\pi}{2})$ and $3\theta^f\in
(0,\frac{\pi}{2})$ then,
$\alpha_1>\beta_1>\beta_3>\alpha_2>\alpha_3>\beta_2.$\\
When, $3\theta^i\in(\pi,\frac{3\pi}{2})$ and $3\theta^f\in
(\frac{3\pi}{2},2\pi)$ then,
$\alpha_1>\beta_1>\beta_2>\alpha_2>\alpha_3>\beta_3.$\\
In all the above cases by equation (2) the states are
incomparable.


\begin{thebibliography}{99}

\bibitem{wootters} W. K. Wootters and W. H. Zurek,  \textit{Nature} {\bf 299}, 802 (1982).
\bibitem{pati11} A. K. Pati and S. L. Braunstein,  \textit{Nature} {\bf 404}, 164 (2000);
W. H. Zurek, \textit{Nature} {\bf 404}, 130 (2000) .
\bibitem{gisin} N. Gisin and S. Popescu,  \textit{Phys. Rev. Lett.} {\bf 83}, 432-435 (1999).
\bibitem{patili} A. K. Pati,  \textit{Phys. Rev. A} {\bf 66},
062319 (2002).
\bibitem{linear} D. Dieks,  \textit{Phys. Lett. A} {\bf 92}, 271 (1982);
H. P. Yuen, \textit{Phys. Lett. A} {\bf 113}, 405 (1986).
\bibitem{mes} A. K. Ekert, \textit{Phys. Rev. Lett.} {\bf 67}, 661 (1991); C. H.
Bennett and S. J. Wiesner, \textit{Phys. Rev. Lett.} {\bf 69},
2881 (1992); C. H. Bennett, G. Brassard, C. Cr\'{e}peau, R. Jozsa,
A. Peres and W. K. Wootters, \textit{Phys. Rev. Lett.} {\bf 70},
1895 (1993).
\bibitem{signal} N. Gisin, \textit{Phys. Lett. A }{\bf 242}, 1-3
(1998); A. K. Pati and S. L. Braunstein, \textit{Phys. Lett. A}
{\bf 315}, 208-212 (2003); A. K. Pati, \textit{Phys. Lett. A} {\bf
270}, 103 (2000).
\bibitem{flip} I. Chattopadhyay, S. K. Choudhary, G. Kar, S. Kunkri and D.
Sarkar, \textit{Phys. Lett. A} {\bf 351}, 384-387 (2006).
\bibitem{horo} M. Horodecki, R. Horodecki, A. Sen(De) and U. Sen, arXiv: quant-ph/0306044.
\bibitem{nielsen} M. A. Nielsen, \textit{Phys. Rev. Lett.} {\bf 83}, 436
(1999); M. A. Nielsen and I. L. Chuang ,`\textit{Quantum
Computation and Quantum Information}' (Cambridge University Press,
2000).
\bibitem{incomparability} I. Chattopadhyay and D. Sarkar,
\textit{Quantum Information and Computation}, {\bf 5}, 247-257
(2005).
\bibitem{unot} V. Buzek, M. Hillery and R. F. Werner, {\it Phys. Rev. A} {\bf 60},
R2626-R2629 (1999); S. J. van Enk, \textit{Phys. Rev. Lett.} {\bf
95}, 010502 (2005).
\bibitem{massar} S. Massar, \textit{Phys. Rev. A }{\bf 62}, 040101(R) (2000).
\bibitem{ghosh} S. Ghosh, A. Roy and U. Sen,  \textit{Phys. Rev. A }{\bf 63}, 014301
(2000); A. K. Pati, \textit{Phys. Rev. A }{\bf 63}, 014302 (2001).
\bibitem{gisin1} C. Simon, V. Buzek and N. Gisin, \textit{Phys. Rev. Lett. }{\bf 87},
170405 (2001).

\end{thebibliography}
\end{document}